\documentclass[aps,prb,reprint,showkeys,showpacs]{revtex4-1}

\usepackage{natbib}
\usepackage{bm}
\usepackage{graphicx}
\usepackage{todonotes}
\usepackage{amsmath}
\usepackage{amsfonts}
\usepackage{xargs} 
\usepackage{color}
\usepackage{epstopdf}
\usepackage{float}
\usepackage[
pdfauthor={derajan},
pdftitle={How to do this},
pdfstartview=XYZ,
bookmarks=true,
colorlinks=true,
linkcolor=blue,
urlcolor=blue,
citecolor=blue,
pdftex,
bookmarks=true,
linktocpage=true, 
hyperindex=true
]{hyperref}
\usepackage{mathtools}

\newcommand\myeq{\stackrel{\mathclap{\normalfont\mbox{\tiny MFA}}}{=}}

\newcommand{\m}{\bm{m}}
\newcommand{\dm}{\frac{d\bm{m}}{dt}}
\newcommand{\HH}{\bm{H}}

\newcommand{\entropic}{\textit{Entropic }}
\newcommand{\magnonic}{\textit{Magnonic }}
\newcommand{\full}{\textit{Full }}

\renewcommand{\r}[1]{{\color{black} #1}}
\renewcommand{\b}[1]{{\color{black} #1}}
\renewcommand{\t}[1]{{\color{black} #1}}

\begin{document}
\title{Domain wall motion by localized temperature gradients}
\author{Simone~Moretti}
\email[Corresponding author: ]{simone.moretti@usal.es}
\author{Victor~Raposo}
\author{Eduardo Martinez}
\author{Luis Lopez-Diaz}
\affiliation{Applied Physics Department, University of Salamanca, Plaza de los Caidos, Salamanca 37008, Spain.}

\begin{abstract}
Magnetic domain wall (DW) motion induced by a localized Gaussian temperature profile is studied in a  Permalloy nanostrip within the framework of the stochastic Landau-Lifshitz-Bloch equation. The different contributions to thermally induced DW motion, entropic torque and magnonic spin transfer torque, are isolated and compared. The analysis of magnonic spin transfer torque includes a description of thermally excited magnons in the sample.  A third driving force due to a thermally  induced dipolar field is found and described.  Finally, thermally induced DW motion is studied under realistic conditions by taking into account the edge roughness. The results give quantitative insights into the different mechanisms responsible for domain wall motion in temperature gradients and allow for comparison with experimental results. 
\end{abstract}


\maketitle

\section{Introduction}
Controlling magnetic domain walls (DW) in ferromagnetic (FM) and antiferromagnetic (AFM) nanostructures  has recently attracted a considerable interest due to its potential for new logic~\cite{Allwood2005} and memory devices~\cite{Parkin2008} and for the very rich physics involved. 
In fact, DWs can be moved by several means such as external magnetic fields~\cite{Schryer1974}, spin polarized currents~\cite{Slonczewski1996, Berger1984,Miron2011,Emori2013} or spin waves~\cite{Han2009, Wang2012,Yan2011,Kim2012}. A new interesting option is the motion of DW by thermal gradients (TG), which was recently observed in few experiments on ferromagnetic (FM) conductors~\cite{Tetienne2014a, Torrejon2012}, semiconductors~\cite{Ramsay2015} and insulators~\cite{Jiang2013}. Spin caloritronics~\cite{Bauer2012} is a new emerging subfield of Spintronics which aims to understand such complex interaction between heat, charge and spin transport. One of the interesting features of thermally induced DW motion is its applicability to FM insulators and AFM~\cite{Selzer2016}. Furthermore, since it does not imply charge transport and related Joule heating, it would avoid energy dissipation in FM conductors or it might represent a solution for harvesting the heat dissipated in electronic circuits.~\cite{Bauer2012, Safranski} 

From a theoretical point of view, it is known that thermally induced DW motion has at least two main causes: (1) the so-called entropic torque (ET)~\cite{Schlickeiser2014, Wang2014a,Kim2015b}, which drives the DW towards the hot region due to maximization (minimization) of Entropy (Free Energy); (2) the magnonic spin transfer torque ($\mu$STT)~\cite{Hinzke2011,Yan2011,Kim2015b}, due to the interaction between thermal magnons, propagating from the hot to the cold region, and the DW. 
While the entropic torque always drives the DW towards the hotter region~\cite{Schlickeiser2014, Wang2014a,Kim2015b,Raposo2016} (the DW energy is always lower where the temperature is higher), the $\mu$STT can drive the DW either towards the hot or the cold part depending on the magnons behavior~\cite{Wang2012}: if magnons are transmitted through the DW, then angular momentum transfer leads to DW motion towards the hot part (opposite the direction of magnon propagation), as predicted in Refs.~\cite{Kim2015b,Hinzke2011,Yan2011}. On the other hand, if magnons are reflected, linear momentum transfer leads to DW motion towards the cold part (the same direction as magnon propagation) as shown in Refs.~\cite{Yan2015,Wang2015a,Han2009}.  Moreover, magnon reflection or transmission depends on many factors such as the DW width, Dzyalonshinskii-Moriya interaction (DMI)~\cite{Wang2015a} and magnon frequency (wavelength)~\cite{Wang2012}. \b{Recently, Kim et al.~\cite{Kim2015c} pointed out another possible mechanism of thermally induced DW motion based on Brownian thermophoresis, which predicts a DW drifting towards the colder region in a thermal gradient. }

As we have briefly described, the picture is rather complex and the main responsible for DW motion in a thermal gradient might depend on the system under investigation. Although numerical studies~\cite{Schlickeiser2014} suggest that the ET is much stronger than $\mu$STT, a detailed comparison is still lacking. Furthermore, previous analyses focused on linear thermal gradients~\cite{Schlickeiser2014, Hinzke2011} where both effects are simultaneously present. However, ET and $\mu$STT have different interaction ranges: the ET is intrinsically \textit{local} (i.e., the DW needs to be inside the TG in order to \textit{feel} the energy gradient and move), while the $\mu$STT  depends on the magnon propagation length~\cite{Ritzmann2014}, which can be larger than the TG extension. Therefore, the dominant effect ($\mu$STT or ET) might depend on the distance from the TG and the comparison between different contributions at different distances remains to be evaluated. Moreover, previous theoretical analysis were performed on perfect samples without considering the effect of pinning, which is essential for comparison with experimental observations.

\begin{figure*}[h]
\centering
\includegraphics[width=0.8\textwidth]{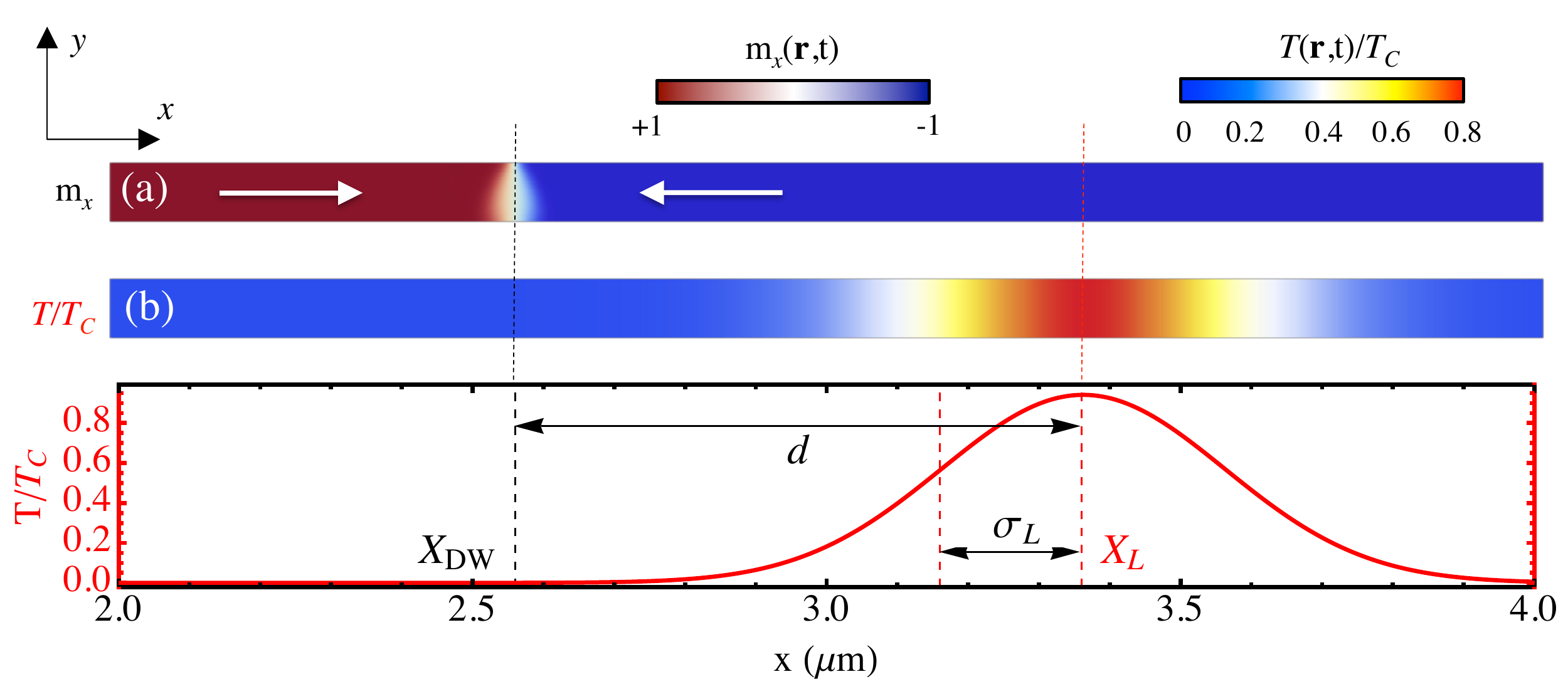}
\caption{(a) Initial magnetization state. (b) Temperature profile along the strip $T/T_C$, with reference to the laser position $X_L$, the  distance $d$ from the DW and the laser width $\sigma_L$.}
\label{fig:Fig1}
\end{figure*}

In this work we study, by means of micromagnetic simulations, the DW motion induced by a \textit{localized} Gaussian temperature profile (as would be given by a laser spot)  placed at different distances from the DW in a Permalloy nanostrip as sketched  in Fig.~\ref{fig:Fig1}. 
We separate magnonic and entropic contributions and we reveal the main responsible for DW motion at each distance. We point out the existence of a third driving force due to a thermally induced dipolar field generated by the laser spot. Such force was ignored before since most theoretical studies were neglecting long-range dipolar interaction.~\cite{Hinzke2011,Schlickeiser2014,Kim2015b} Finally, by including edge roughness, we analyse the thermally induced DW motion under realistic conditions. The article is structured as follows: Sec.~\ref{secII} describes the numerical methods and the system under investigation. The main observations are outlined in Sec.~\ref{secIII} while the different driving mechanism are explained in more details in Sec.~\ref{subsec1} (Entropic torque), \ref{subsec2} (Thermally induced dipolar field) and \ref{subsec3} (Magnonic spin transfer torque). Finally, the results for a realistic strip are shown in Sec.~\ref{subsec4} and the main conclusions are summarized in Sec.~\ref{sec4}.\\

\section{Methods}
\label{secII}
Magnetization dynamics is analysed in a Permalloy nanostrip of length $L_x=5.12 {\rm \mu m}$ and cross section $S=(80\times 10){\rm nm^2}$ with a head-to-head transverse DW (TW) placed and relaxed in its center ($X_{DW}=2.56 {\rm \mu m}$). The initial magnetic configuration and reference frame are shown in Fig.~\ref{fig:Fig1}(a). Magnetization lies in-plane along the $x$ direction and the TW is stable for these dimensions. Magnetic evolution is studied by means of the stochastic Landau-Lifshitz-Bloch (LLB) equation:~\cite{Garanin1997,Chubykalo-Fesenko2006,Kazantseva2008, Evans2012,Moretti2016}

\begin{eqnarray}
\dm&=&-\gamma_0(\m\times\HH_{\rm{eff}})+\frac{\gamma_0\alpha_{\parallel}}{m^2}(\m\cdot\HH_{\rm{eff}})\m\nonumber\\&&-\frac{\gamma_0\alpha_{\perp}}{m^2}[\m\times(\m\times\HH_{\rm eff}+\bm{\eta}_{\perp})]+\bm{\eta}_{\parallel}\, ,
\label{eq:LLB}
\end{eqnarray}
where $\gamma_0$ is the gyromagnetic ratio. $\alpha_{\perp}=\alpha_0[1-T/(3T_C)]$ and $\alpha_{\parallel}=\alpha_0(2T/3T_C)$ are the transverse and longitudinal damping parameters respectively, where $\alpha_0$ is a microscopic damping parameter coupling the spins to the lattice, and $T_C$ indicates the Curie temperature. $\m$ represents the normalized magnetization vector ($\m=\bm{M}/M_0$, $M_0$ being the saturation magnetization at zero temperature) and $\HH_{\rm eff}$ the effective magnetic field given by: 
\begin{eqnarray}
\HH_{\rm eff}=\frac{2A(T)}{\mu_0M_0m_e^2}\nabla^2\m+\HH_{\rm dmg}+\frac{1}{2\tilde{\chi}_{\parallel}}\left(1-\frac{m^2}{m_e^2}\right)\m\, .\nonumber\\
\label{eq:Heff}
\end{eqnarray}
The first term on the right-hand side (RHS) is the exchange field~\cite{Schlickeiser2014, Hinzke2011} ($A(T)$ is the temperature dependent exchange stiffness, $\mu_0$ is the vacuum permeability, and $m_e$ is the equilibrium magnetization module). $\HH_{\rm dmg}$ is the demagnetizing field, while the last term represents the longitudinal exchange field, which drives the module of $\m$ towards its equilibrium value at each temperature, $m_e(T)$. $\tilde{\chi_{\parallel}}$ is the longitudinal susceptibility defined as $\tilde{\chi_{\parallel}}=(\partial m_e/\partial H_a)_{H_a\rightarrow 0}$, with $H_a$ being the external field. The choice of LLB is preferred over the conventional Landau-Lifshitz-Gilbert (LLG) equation since it  allows us to describe magnetization dynamics for temperatures even close to $T_C$. Furthermore, it naturally includes the ET within the temperature dependence of the micromagnetic parameters $\m (T)$ and $A(T)$~\cite{Schlickeiser2014}. In fact, in the LLB, $\m$ is not restricted to unity and its value depends on the temperature, as well as the values of $\tilde{\chi_{\parallel}}$, $\alpha_{\perp}$ and $\alpha_{\parallel}$. However, the ET only depends on $\m (T)$ and $A(T)$, since they directly affect the DW energy, while the other parameters ($\tilde{\chi_{\parallel}}$, $\alpha_{\perp}$, $\alpha_{\parallel}$) affect the dynamics.

\begin{figure}[H]
\centering
\includegraphics[width=0.4\textwidth]{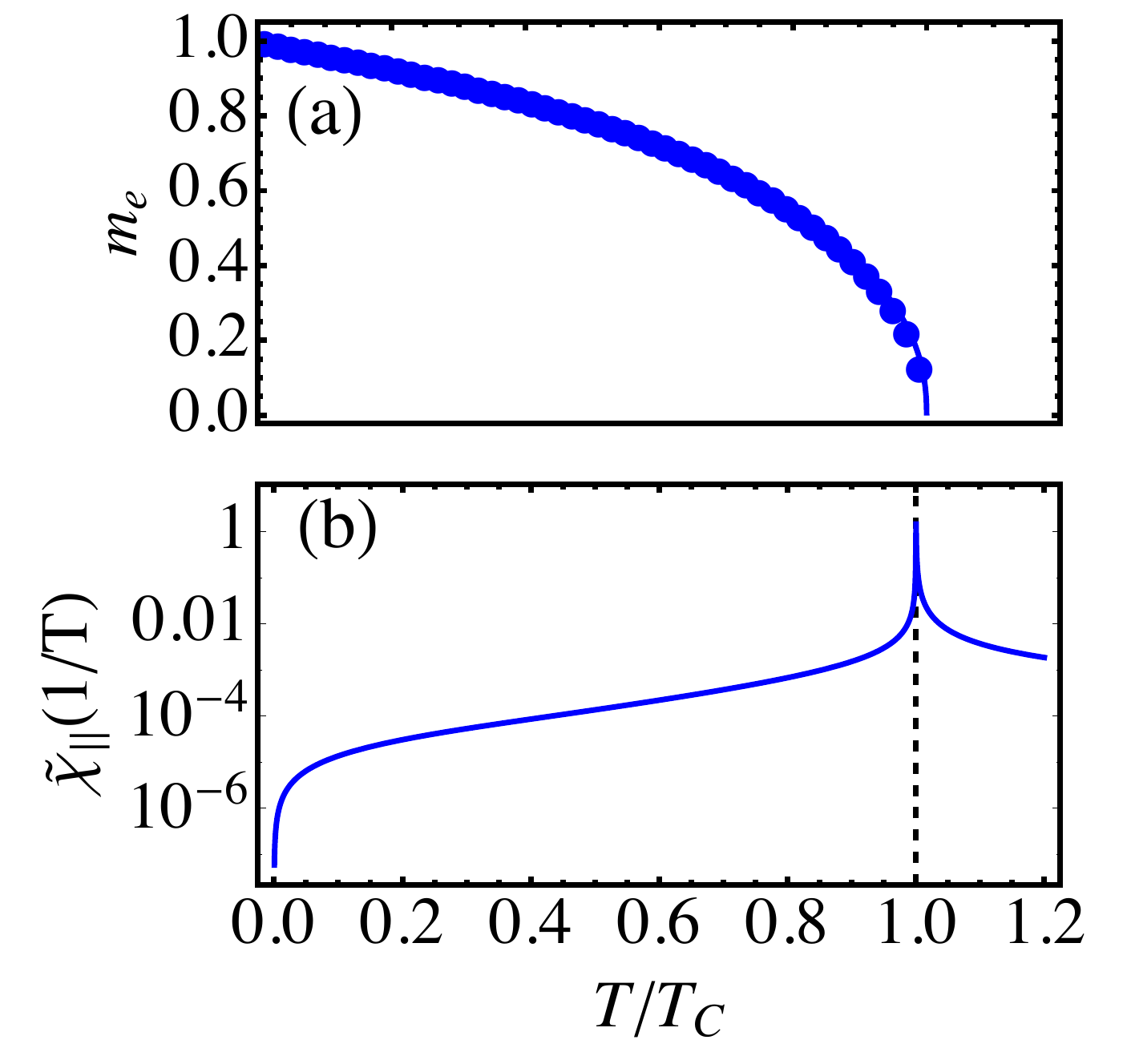}
\caption{\r{(a) Equilibrium magnetization $m_e(T)$ obtained with the Langevin function~\cite{Bertotti}. Dots correspond to numerical results while the solid line is a fit of the numerical solution. (b) Longitudinal susceptibility $\tilde{\chi_{\parallel}}=(\partial m_e/\partial H_a)_{H_a\rightarrow 0}$.}}
\label{fig:me}
\end{figure}

The function $m_e(T)$ needs to be introduced as an input into the model and, within the mean-field approximation (MFA) and in the classic limit, it is given by the Langevin function:~\cite{Bertotti,Atxitia2007} $m_e=\coth(x)-1/x$, \r{with $x=\mu_0\mu_{\rm Py}H_a/(k_B T)+3T_Cm_e/T$, where $\mu_{\rm Py}$ is the Py magnetic moment and $k_B$ the Boltzmann constant. The second term represents the Weiss molecular field expressed in term of $T_C$. For the calculation we considered $H_a\rightarrow 0$ since there is no applied field. The obtained $m_e$ and $\tilde{\chi_{\parallel}}$ are shown in Fig.~\ref{fig:me}. } Within the MFA the temperature dependence of the exchange stiffness is given by $A(T)=A_0m_e(T)^2$, ~\cite{Atxitia2007,Ramsay2015,Moretti2016} where $A_0$ is the exchange stiffness at $T=0$.  The dynamics of the magnetization module (the \textit{longitudinal} dynamics) is described by the second term on the RHS of Eq.~(\ref{eq:LLB}), proportional to $\alpha_{\parallel}$ and it is governed by the longitudinal exchange field in Eq.~(\ref{eq:Heff}), proportional to $\tilde{\chi}_{\parallel}$. Such dynamics becomes important at $T\approx T_C$ when longitudinal and transverse relaxation times are comparable. 

$\bm{\eta}_{\perp}$ and $\bm{\eta}_{\parallel}$ are transverse and longitudinal stochastic fields, which introduce thermal fluctuations - and therefore excite thermal magnons - into the system. They have white noise properties with correlators given by~\cite{Evans2012}
\begin{eqnarray}
\left\langle \eta_{\perp}^i(\bm{0},0)\eta_{\perp}^{j}(\bm{r},t)\right\rangle&=&\frac{2k_BT(\alpha_{\perp}-\alpha_{\|})}{\gamma_0\mu_0M_s^0V\alpha_{\perp}^2}\delta_{ij}\delta(\bm{r})\delta(t)\, ,\nonumber\\
\left\langle \eta_{\parallel}^i(\bm{0},0)\eta_{\parallel}^{j}(\bm{r},t)\right\rangle&=&\frac{2\gamma_0k_BT\alpha_{\|}}{\mu_0M_s^0V}\delta_{ij}\delta(\bm{r})\delta(t)\, ,
\label{eq:noise}
\end{eqnarray}
which are obtained by imposing the Maxwell-Boltzmann distribution as the solution of the Fokker-Plank equation calculated from the stochastic LLB~\cite{Evans2012}. \r{Note that a cut-off on the magnon wavelength is imposed by discretizing the sample in cubic cells (as commonly done in finite-difference solvers); that is,  magnons with wavelength smaller than $2\Delta x$ cannot be included in the thermal noise, where $\Delta x$ is the cell size. Within the LLB formalism, their contribution is still included in the temperature dependence of the micromagnetic parameters $m_e(T)$ and $A(T)$, but any flux of such small wavelength magnons, from one cell to another, is neglected.  This means that our analysis of the magnonic STT is restricted to thermal magnons with $\lambda > 2\Delta x$. For this reason we chose cells of dimensions $2.5\times 5\times 10{\rm nm}$  in order to include a higher flux of magnons along the $x$ direction.   
In Sec.~\ref{subsec3} we will see that these magnons have a very small propagation length ($L_p\sim80$nm, see Eq.~\ref{eq:lp}) and therefore they can be ignored when they are excited far from the DW position, since they do not reach it. Inside the TG, where they could reach the DW, their contribution is ignored and this constitutes a limitation of our model.} \t{At the end of Sec.~\ref{subsec3} we will briefly discuss the possible effects of such small wavelength magnons. }

In summary, the ET is naturally included into the model by the temperature dependence of $\m(T)$ and $A(T)$ while magnons are excited by the stochastic fields. At a given temperature we would have both effects simultaneously. To isolate the effect of the ET we simply perform simulations without thermal field and we  label these kind of simulations as  \textit{Entropic}. To isolate the effect of magnons we perform simulations by keeping $\m (T)$ and $A(T)$ constant at their $T=0$ values and we label these kind of simulations as \textit{Magnonic}. Simulations with the full stochastic LLB (Eq.~(\ref{eq:LLB})) are labelled as \textit{Full}. The \magnonic simulations correspond to what one would observe within the LLG framework for $T\ll T_C$, assuming that $m_e(T)$ and $A(T)$ do not change with temperature. Indeed we checked that for $T_L=200,400$K (where the LLG framework can be applied) the \magnonic results correspond with the results of the conventional LLG. 

Eq.~(\ref{eq:LLB}) is solved by finite difference method with a customized software.~\cite{Moretti2016,Raposo2016} 
We use the mentioned cell size ($2.5\times5\times10\ {\rm nm}$) and a time step of $0.1{\rm ps}$ testing that smaller time steps produce equal results. Typical Py parameters are considered: $A_0=1.3\times 10^{-11} {\rm J/m}$, $M_0=8.6\times10^5\ {\rm A/m}$, $\alpha_0=0.02$ and $T_C=850$K.

The strip temperature is given by a Gaussian profile:
\begin{eqnarray}
T(x)=T_0+T_L \exp\left[{-\frac{(x-X_L)^2}{2\sigma_L^2}}\right]\, ,
\label{eq:Tprofile}
\end{eqnarray}
where $T_0=0$ and $T_L$ is the laser temperature. $X_L$ is the laser spot position, and $\sigma_L$ is the Gaussian profile width. For our study we chose $\sigma_L=200 {\rm nm}$, which would correspond to a laser waist of $\sqrt{2}\sigma_L\approx 280$ nm, reasonable for typical lasers.~\cite{Tetienne2014a} We performed simulations placing the laser spot at different distances from the DW ($d=X_L-X_{DW}$). Distances correspond to integer multiples of $\sigma_L$  i.e. $d=X_L-X_{DW}=N\sigma_L$ with $N=1,2,...,10$. Simulations are performed for $4$ different laser temperature $T_L=200,400,600$ and $800K$. The temperature profile is plotted in Fig.~\ref{fig:Fig1}(b) for $T_L=800K$. Five different stochastic realizations are considered when thermal fluctuations are taken into account (\magnonic and \full simulations). The gradient extension from $X_L$ is approximatively equal to $3\sigma_L$, in other words  $\nabla T(X_{\rm DW})\approx 0$ if $d>3\sigma_L$. In fact, $\nabla T(3\sigma_L)\approx 0.1$K/nm and the estimated entropic field (see Sec.~\ref{subsec1}) is $\mu_0H_E\approx 0.06$ mT. Furthermore, $T=0$K is imposed for $d>4.5\sigma_L$ so that $\nabla T(X_{\rm DW})= \mu_0H_E = 0$ if $d>4.5\sigma_L$.
We simulate an infinite strip by removing the magnetic charges appearing at both sides of the computational region~\cite{Martinez2007a}. 
The simulation time window is varied depending on  $d$ and the DW velocity, with a maximum simulation time of $t_{max}=500{\rm ns}$.

\section{Results and Discussion}
\label{secIII}

\begin{figure*}[h]
\centering
\includegraphics[width=1.0\textwidth]{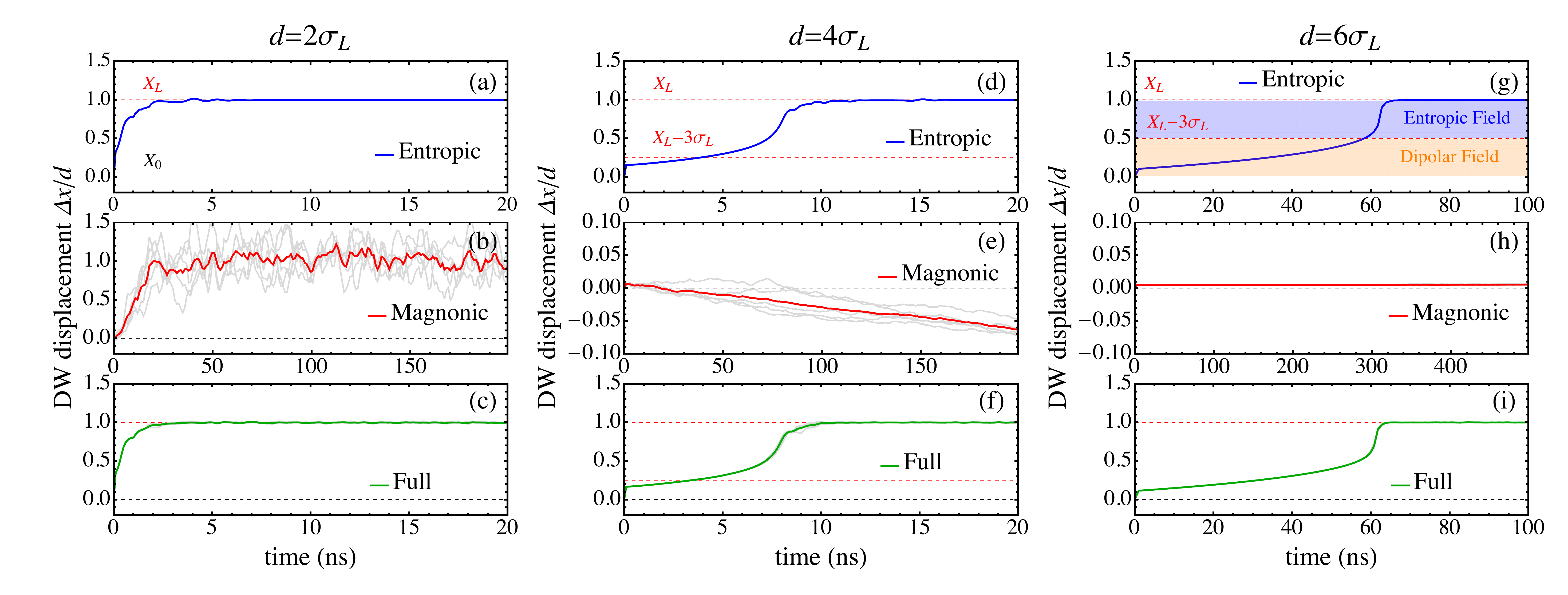}
\caption{DW displacment as function of time for $d=2\sigma_L$ (a,b,c ), $d=4\sigma_L$ (d,e,f) and $d=6\sigma_L$ (g,h,i), for the \entropic, \magnonic and \full cases as labelled in each plots. Displacement $\Delta x$ is normalized to $d$: a displacement of $1$ means that the DW has reached the center of the laser spot. $X_L$ indicates the laser position, $X_0$ the DW position and $X_L-3\sigma_L$ the extension of the TG i.e. the region where $\nabla T(x)\neq 0$.}
\label{fig:Fig2}
\end{figure*}

Fig.~\ref{fig:Fig2} shows the normalized DW displacement ($\Delta x/d$) as function of time for the \entropic, \magnonic  and \full cases, calculated with $T_L=800K$, for three different distances as labelled on top of each column: $d=2\sigma_L$ (Fig.~\ref{fig:Fig2}(a)-(c))~,  $d=4\sigma_L$(Fig.~\ref{fig:Fig2}(d)-(f))  and  $d=6\sigma_L$ \ref{fig:Fig2}(g)-(i)). The displacement is normalized to the laser distance $d$ and therefore $\Delta x/d=1$ indicates that the DW has reached the laser spot.

For $d=2\sigma_L$ the DW moves towards the laser spot both for \magnonic (Fig.~\ref{fig:Fig2}(b)) and \entropic (Fig.~\ref{fig:Fig2}(a)) cases (and therefore obviously in the \full (Fig.~\ref{fig:Fig2}(c)) case). The DW is inside the TG ($\nabla T(X_{\rm DW})\neq 0$) and  its motion can be attributed mostly to the  ET.~\cite{Schlickeiser2014,Wang2014a,Kim2015b} In the \magnonic case the motion could be given by a magnons stream passing adiabatically through the DW, but also by the effect of an averaged ET. Note that  thermal magnons, apart from the $\mu STT$, also introduce an averaged ET~\cite{Kim2015b,Raposo2016}: where the temperature is higher, the \textit{averaged} $M_s$ (over more cells) is lower (as in the \entropic case) due to higher thermal fluctuation. More precisely, we recall that also the temperature dependence of $\m$ and $A$ is given by averaged high frequency magnons which cannot be included in the thermal fluctuations due to the spatial discretization.  This averaged ET is, however, a small contribution compared to the one given by high frequency magnons as can be seen by the time scale in Fig.~\ref{fig:Fig2}(a) and (b). 

For $d=4\sigma_L$ the DW moves towards the hotter region in the \entropic  case (Fig.~\ref{fig:Fig2}(d)) and towards the colder region in the \magnonic case (Fig.~\ref{fig:Fig2}(e)). The latter could be attributed to $\mu$STT (Sec.~\ref{subsec3}) while the first effect is unexpected since at $d=4\sigma_L$, $\nabla T(X_{\rm DW})\approx 0$ and the ET should have no effect (we recall that the gradient extension is approximatively $3\sigma_L$). This is even more visible in the $d=6\sigma_L$ case  where, although it is certain that $\nabla T(X_{\rm DW})=0$, the DW moves towards the hotter region (Fig.~\ref{fig:Fig2}(g)). Note that the DW moves with different velocities when it is outside ($x>X_L-3\sigma_L$) or inside ($x<X_L-3\sigma_L$) the TG. We conclude that there must be another long-range driving force - not related to magnons - that drives the DW in this case. As we will discuss in Sec.~\ref{subsec2} this force is given by a thermally induced dipolar field. In the \magnonic case (Fig.~\ref{fig:Fig2}(h)) the DW does not move, compatibly with $\mu$STT if we assume that magnons are already damped for such distance ($d=6\sigma_L=1.2\ {\rm \mu m}$). Indeed we estimated a magnon propagation length of $330\ {\rm nm}$ ($\ll 1.2 {\rm \mu m}$)  for our sample (Sec.~\ref{subsec3}). Different laser temperatures ($T_L=200,400,600K$) produce qualitatively similar behaviors for all the distances. Furthermore, for all cases the \full simulations are very similar to the \entropic simulations  suggesting that the ET dominates over the $\mu$STT.~\footnote{To be sure that the results are not numerical artefacts we performed simulations for the symmetric cases $d=-2,-4,-6\sigma_L$ obtaining equal results.}

\begin{figure}[H]
\includegraphics[width=0.4\textwidth]{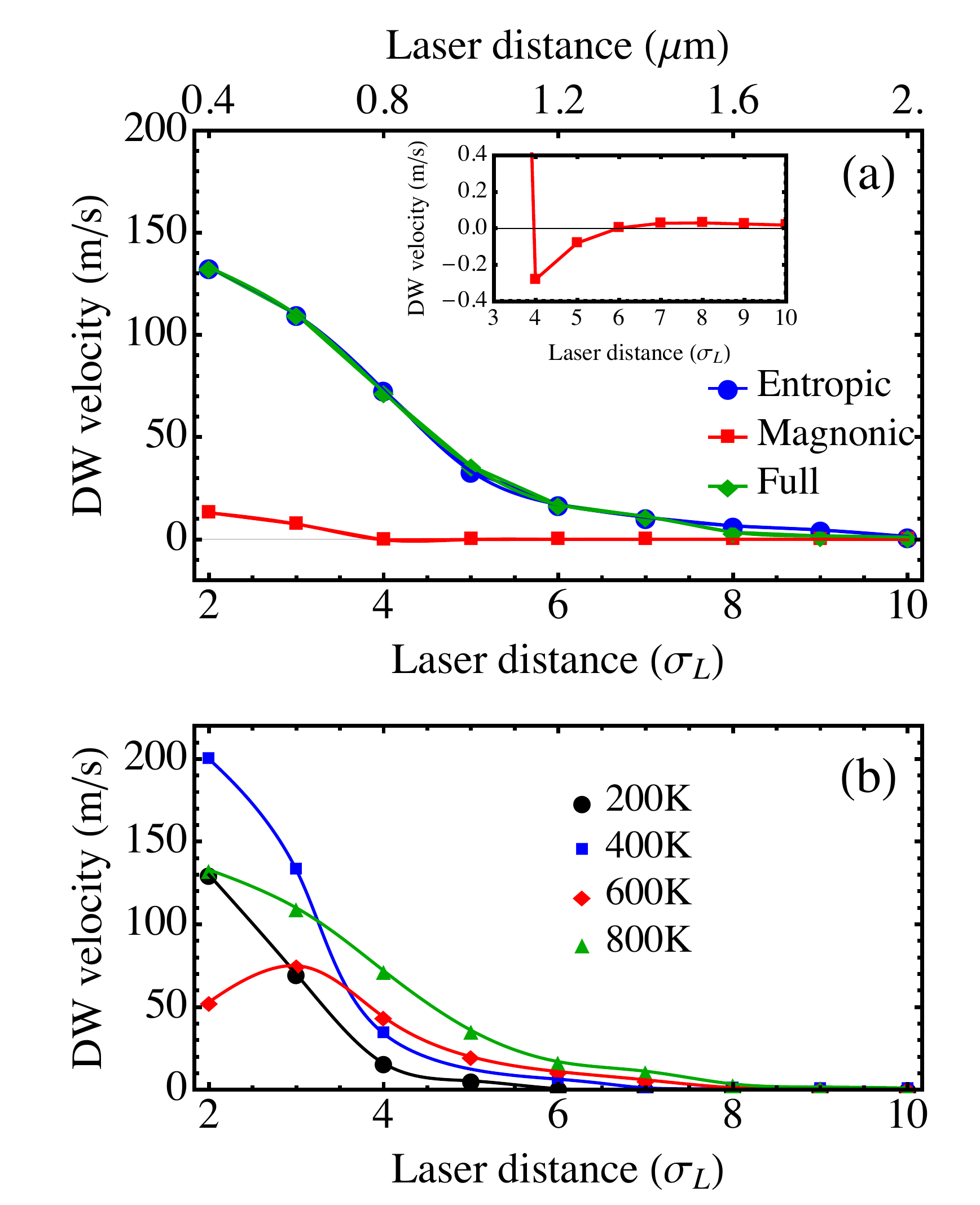}
\caption{(a) DW average velocity as function of laser distance, for the \entropic (Blue dots), \magnonic (Red squares) and \full cases (Green diamonds) respectively. The inset shows a detail of the \magnonic case from $d=4\sigma_L$ to show the negative small velocities, not visible in comparison with the \entropic velocities. (b) DW average \full velocities for different temperatures. At $d=2\sigma_L$ the maximum velocity is observed at $T=400$K due to the WB threshold at $T\geq600$K. }
\label{fig:Fig3}
\end{figure}

Fig~\ref{fig:Fig3}(a) shows the averaged DW velocity~\footnote{If the DW reaches the laser spot, the averaged velocity is obtained as $d/t_L$ where $t_L$ is the time needed to reach the spot. If the DW does not reach the laser spot, then the velocity is obtained as $d/t_{max}$, where $t_{max}$ is the maximum simulation time.} as function of laser distance. As mentioned, the ET dominates the DW dynamics and in fact \entropic and \full velocities almost coincide. The averaged velocity decreases with distance but it is different from $0$ even for the maximum distance we analysed ($d=10\sigma_L=2\ {\rm \mu m}$) meaning that, with enough time $t>t_{max}$, the DW would reach the laser spot since the velocities are always positive (towards the hot part). On the other hand, \magnonic simulations give rise to positive velocity for $d=2,3\sigma_L$ due to an averaged ET, negative velocities (towards the cold part) for $d=4,5\sigma_L$ and almost null velocities for $d\ge 6\sigma_L$. In all the cases, the velocities due to $\mu$STT are much smaller than the \entropic velocities, in agreement with previous predictions~\cite{Schlickeiser2014}. Details of the averaged magnons velocities are shown in the inset of Fig.~\ref{fig:Fig3}(a)\\

Fig.~\ref{fig:Fig3}(b) displays the average \full velocity as function of laser distance for  $4$ different laser temperatures: at $d=2,3\sigma_L$ the maximum DW velocity is observed for $T_L=400$K due to the Walker breakdown (WB) threshold at $T_L=600$K, predicted also for thermal induced DW motion~\cite{Schlickeiser2014}: Below $600$K the entropic field (Sec.~\ref{subsec1}) is compensated by the DW shape anisotropy and the DW moves rigidly without changing its internal structure. For $T_L>600K$ the DW anisotropy cannot compensate the entropic field and the DW precesses, changing its internal structure and resulting in a slower velocity, as can be seen in a movie in the Supplemental Material~\footnote{see Supplementary Material at URL for movies of the Walker breakdown mechanism and realistic sample depinning.}. On the other hand, for $d\ge 4\sigma_L$ the maximum temperature coincides with maximum velocity since the thermally induced dipolar field is below the WB and the DW moves faster while outside the TG. In what follows we analyse each contribution separately.

\subsection{Entropic field}
\label{subsec1}
The ET originates from the fact the the DW free energy ($\Delta F(T)$) decreases with temperature and, as a consequence, the DW moves towards the hotter region in order  to minimize its free energy.~\cite{Hinzke2011,Schlickeiser2014,Wang2014a} It is called \entropic since DW entropy increases with temperature and leads to the overall decrease of the free energy~\cite{Hinzke2008,Wang2014a,Schlickeiser2014}, $\Delta F=\Delta U-T\Delta S$, with $\Delta U$ being the DW internal energy~\cite{Hinzke2008,Schlickeiser2014} and $\Delta S$ being the DW entropy~\cite{Hinzke2008,Schlickeiser2014}. In the thermodynamic  picture of LLB, entropy is included in the temperature dependent DW 
free energy density~\cite{Hinzke2008,Schlickeiser2014,Hillebrands3}
\begin{eqnarray}
\epsilon_{\rm DW}(T)=4\sqrt{A(T)(K_0(T)+K_S(T)\sin^2\phi)}\, ,
\end{eqnarray}
where $K_0(T)$ and $K_S$ are effective anisotropy constants, and $\phi$ is the internal DW angle. In the case of Permalloy, the anisotropies are both of magnetostatic origin (shape anisotropies) and they are given by
\begin{eqnarray}
K_0(T)&=&\frac{1}{2}\mu_0M_0^2(N_y-N_x)m(T)^2\, ,\nonumber\\
K_S(T)&=&\frac{1}{2}\mu_0M_0^2(N_z-N_y)m(T)^2\, ,
\end{eqnarray}
being $N_{x,y,z}$ the demagnetizing factors. As in MFA also $A(T)$ decreases with $T$ as $m(T)^2$, $\epsilon_{\rm DW}(T)$ decreases as
\begin{eqnarray}
\epsilon_{\rm DW}(T)=4\sqrt{A_0(K_0^0+K_S^0\sin^2\phi)}m(T)^2\, ,
\label{eq:dw_energy}
\end{eqnarray}
where  $K_0^0$ and $K_S^0$ are the shape anisotropies at $T=0$.
Therefore, the temperature gradient introduces a DW energy gradient, which leads to the equivalent field (the so-called \entropic field)
\begin{eqnarray}
\mu_0\HH_E&=&-\frac{1}{2m_eM_0}{\bf \nabla}\epsilon_{\rm DW}=-\frac{1}{2m_eM_0}\frac{\partial\epsilon_{\rm DW}}{\partial T}\frac{\partial T}{\partial x}\hat{x}\, \nonumber\\
&=&-\frac{4A_0}{M_0\Delta_0}\frac{\partial m}{\partial T}\frac{\partial T}{\partial x}\hat{x}\, ,
\label{eq:ET}
\end{eqnarray}
where the gradient is only along $\hat{x}$ and
\begin{eqnarray}
\frac{\partial \epsilon_{\rm DW}(T)}{\partial T}&=&2m\epsilon_{\rm DW}^0\frac{\partial m}{\partial T}
=\frac{8mA_0}{\Delta_0}\frac{\partial m}{\partial T}\, .
\end{eqnarray} 
$\Delta_0=\sqrt{A_0/(K_0+K_S^0\sin^2\phi)}$ and $\epsilon_{\rm DW}^0$ are the DW width and energy at $T=0$ respectively.  

In Ref.~\cite{Schlickeiser2014} Schlickeiser et al. proposed an analytical expression for $\mu_0 H_E^{*}$ by solving the LLB equation in the 1D approximation. Within the MFA their expression is indeed equivalent to Eq.~(\ref{eq:ET}). In fact,
\begin{eqnarray}
\mu_0\HH_E^{*}&=&-\frac{2}{\Delta (T) M_0}\frac{\partial A(T)}{\partial T}\frac{\partial T}{\partial x}\hat{x}\nonumber\\
&\myeq&-\frac{4A_0}{\Delta_0 M_s}\frac{\partial m}{\partial T}\frac{\partial T}{\partial x}\hat{x}\, .
\end{eqnarray} 

\begin{figure}[H]
\centering
\includegraphics[width=0.45\textwidth]{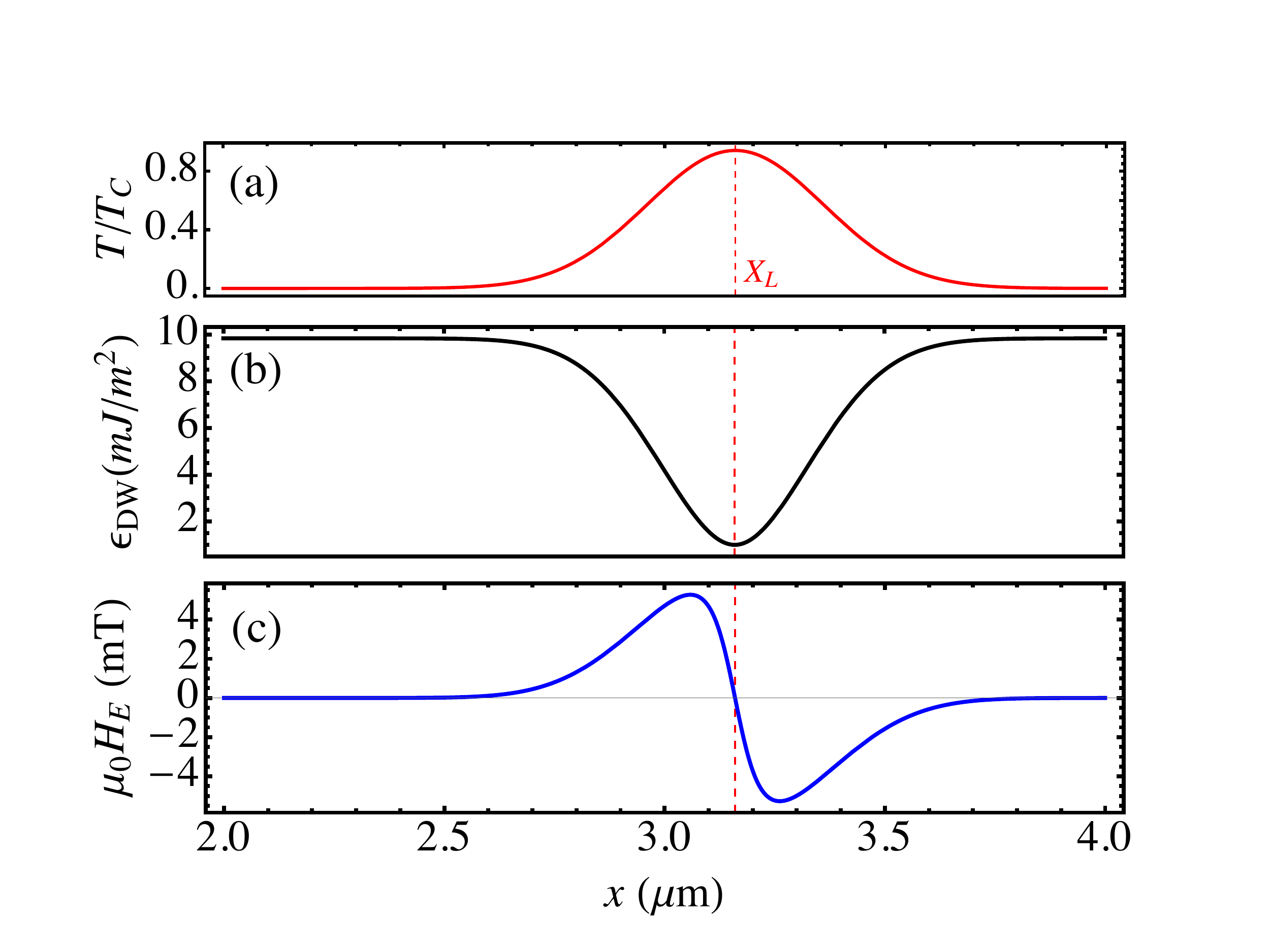}
\caption{(a) Temperature profile $T(x)$ for $X_L=3.16{\rm \mu m}$ and $T_L=800$K. (b) Corresponding DW energy profile $\epsilon_{DW} (T(x))$ as given by Eq.~(\ref{eq:dw_energy}) and (c) the resulting Entropic field as predicted by Eq.~(\ref{eq:ET}).}
\label{fig:Fig4}
\end{figure}

\b{Fig.~\ref{fig:Fig4}(a) shows the strip temperature profile for $T_L~=~800$K and $X_L=3.16{\rm \mu m}$.  Fig.~\ref{fig:Fig4}(b) depicts the corresponding DW energy profile $\epsilon_{\rm DW}(T(x))$ (Eq.~(\ref{eq:dw_energy})) and Fig.~\ref{fig:Fig4}(c) the resulting entropic field $\mu_0H_E(x)$ (Eq.~(\ref{eq:ET})). The entropic field always pushes the DW towards the center of the laser spot where $\mu_0H_E=0$ since $\nabla T=0$. Note that indeed, the ET is \textit{local} because it depends on $\nabla T(x)$: if $\nabla T(x)=0$ then $\mu_0H_E(x)=0$.  The maximum field is approximatively at $d=1\sigma_L$, where $\nabla T$ is maximum. }

\subsection{Thermally induced dipolar field}
\label{subsec2}
\begin{figure}[H]
\centering
\includegraphics[width=0.45\textwidth]{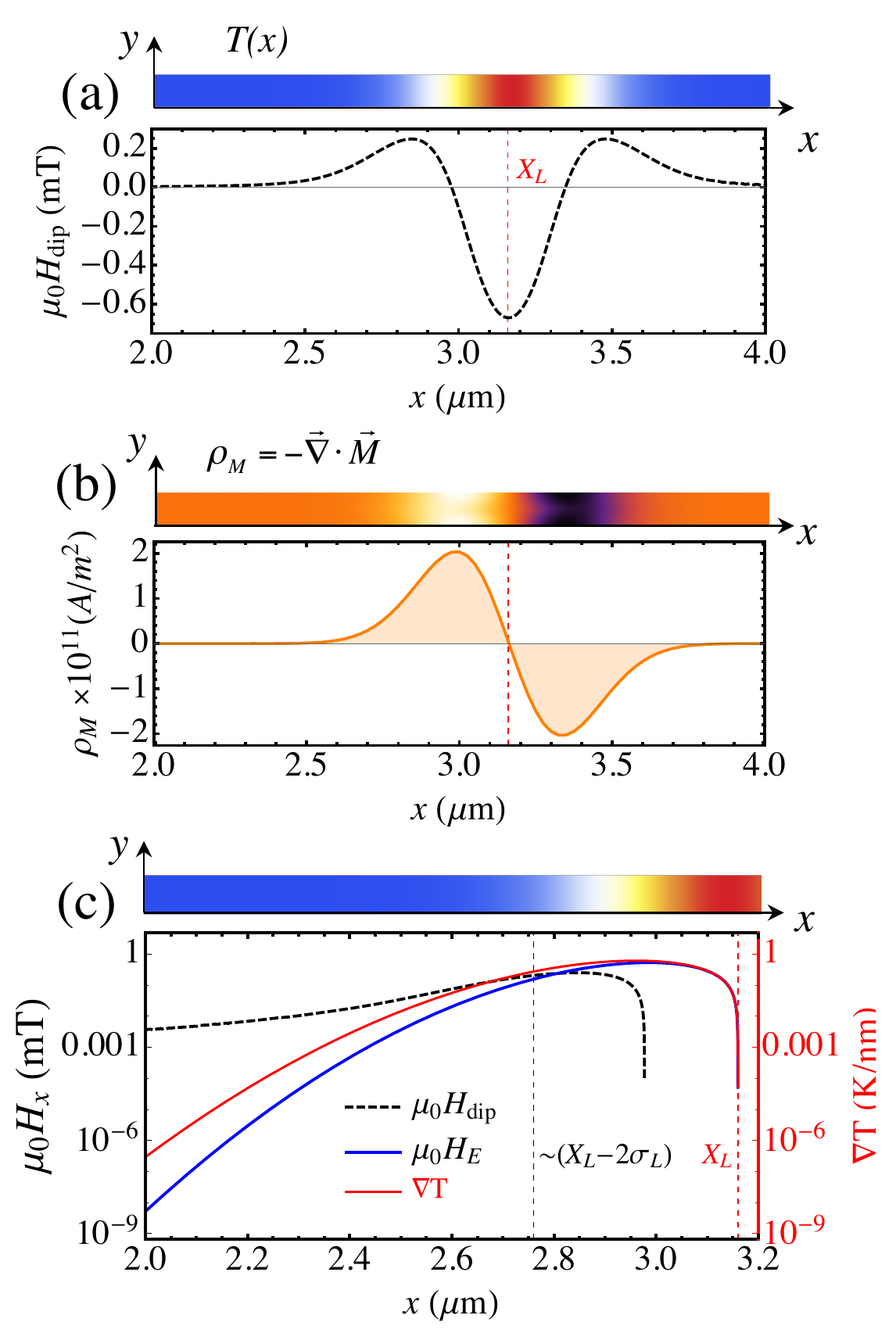}
\caption{(a) Thermally induced dipolar field and (b) volume charges $\rho_M=-\nabla \cdot \bf{M}$  for $X_L=3.16{\rm \mu m}$ and $T_L=200$K. (c) Comparison between the entropic and demagnetizing field. Beyond $2\sigma_L$ from the laser spot, the demagnetizing field dominates.}
\label{fig:Fig5}
\end{figure}

Since in the \entropic simulations the DW moves  even when $\nabla T(X_{\rm DW})=0$, there must be another force responsible for its motion at large distances. A natural candidate is the demagnetizing field which is a long-range interaction. Indeed, a thermally induced dipolar field (TIDF) was found to be the responsible for the DW motion at large distances. Fig.~\ref{fig:Fig5}(a) displays the TIDF ($H_{\rm dip}$) of a uniform magnetized strip with the laser spot. Strip magnetization is saturated along $x$ ($m_x=-1$) and the TIDF is calculated by subtracting the demagnetizing field of the strip without the laser spot from the demagnetizing field of the same uniform strip with the laser spot (in this way we can isolate the effect of the laser). The field has a minimum at $X_L$ and positive tails outside the thermal gradient (Fig.~\ref{fig:Fig5}(a),(c)). The laser temperature is set to the minimum value $T_L=200$K. The TIDF is due to the volume charges $\rho_M=-\nabla\cdot\bf{M}$, shown in Fig.~\ref{fig:Fig5}(b), which arise from the variation of magnetization module. Positive and negative charges, on the left and right side of $X_L$ respectively, sum their effect in the center giving rise to the minimum value of the TIDF (maximum in module) while they compete each other outside the laser spot giving rise to the decaying behavior. 

A comparison between the TIDF and the entropic field is shown in Fig.~\ref{fig:Fig5}(c). As expected, beyond $2\sigma_L$ the TIDF is  much larger than the entropic field that rapidly decays to $0$ outside the TG. $\mu_0H_E$ decays as $\nabla T\propto (x-X_L)e^{-|x-x_L|^2/(2\sigma_L^2)}$ as expected, while the TIDF decays as $1/x^3$ as expected for a dipolar field (Fig.~\ref{fig:Fig5}(c)). Before $2\sigma_L$ the comparison has no meaning since the TIDF is calculated for a uniform magnetization and it would change once the DW approaches the laser center. 

\begin{figure}[H]
\centering
\includegraphics[width=0.4\textwidth]{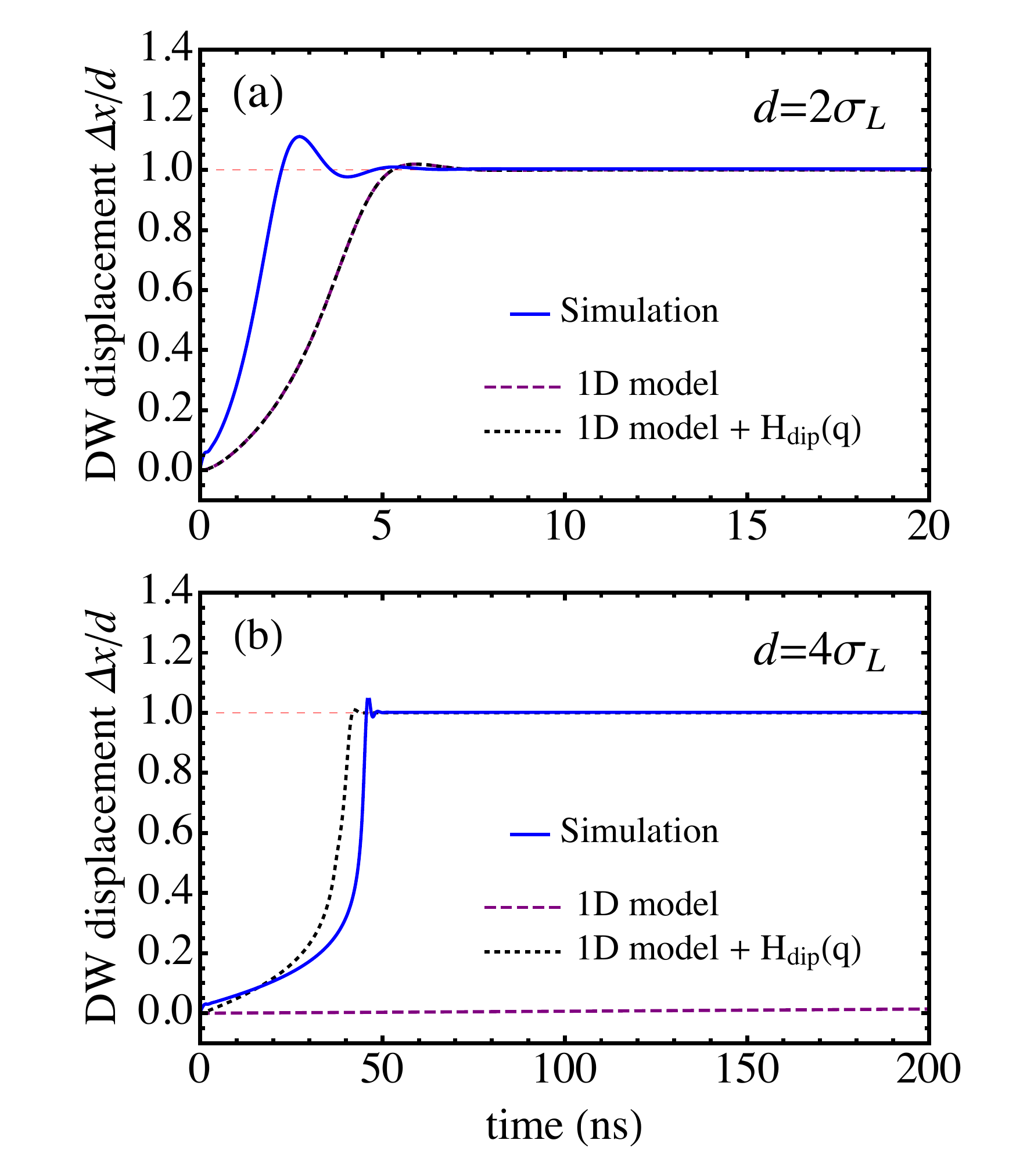}
\caption{DW displacement as function of time as predicted by the 1D model with or without the demagnetizing field for $d=2\sigma_L$ (a) and $d=4\sigma_L$ (b). The model without demagnetizing field does not predict any motion for $d\geq4\sigma_L$. }
\label{fig:Fig6}
\end{figure}

\begin{figure}[H]
\centering
\includegraphics[width=0.38\textwidth]{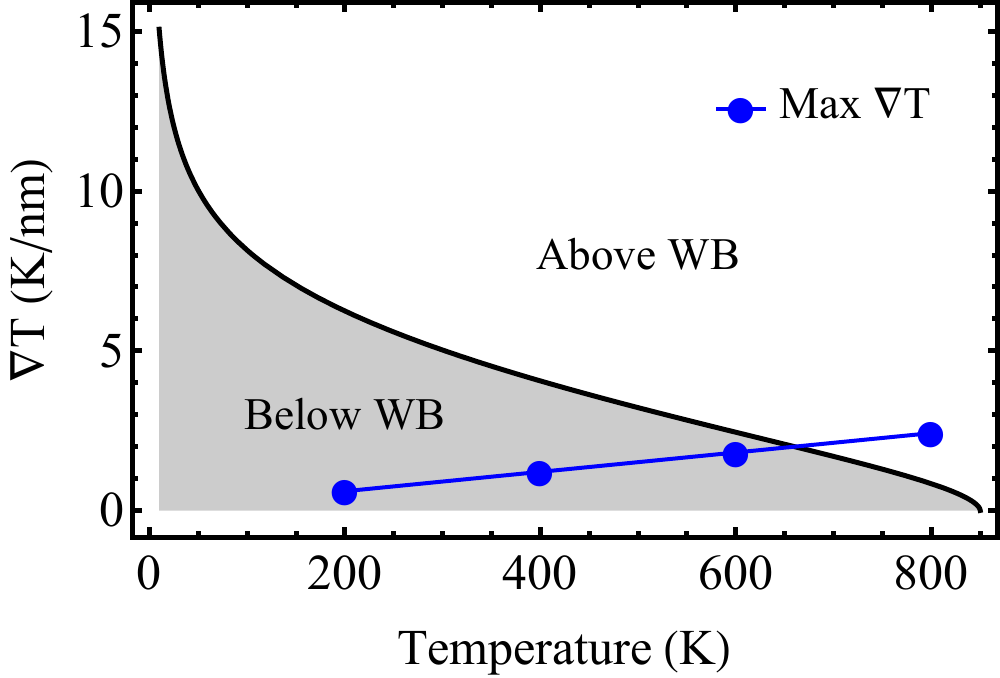}
\caption{WB thermal gradient as function of temperature (Eq.~(\ref{eq:WB})). For a Gaussian profile, as the one applied in our simulations, $\nabla T$ is not constant, the blue points represent the maximum value of $\nabla T$ for each temperature which occurs approximatively at $1\sigma_L$ from the laser spot center. }
\label{fig:Fig6bis}
\end{figure}

To further check our explanation, a 1D model  was implemented following Ref.~\cite{Schlickeiser2014}. The model originally includes the ET while the TIDF was added by fitting the micromagnetic TIDF (Fig.~\ref{fig:Fig5}(a)). The field is set different from zero only if $d>2\sigma_L$ since it has no meaning for closer distances as previously commented.  The 1D model equations governing the DW internal angle $\phi$ and DW position $q$ read like:
\begin{eqnarray}
\dot{\phi}&=&\gamma_0\left[\left(H_{\rm dip}-\frac{4A_0}{\mu_0 M_0\Delta_0}\frac{\partial m}{\partial x}\right)-\alpha_{\perp} \frac{K_S^0}{\mu_0 M_0}\sin(2\phi) \right]\nonumber\, ,\\
m\frac{\dot{q}}{\Delta_0}&=&\gamma_0\left[\alpha_{\perp}\left(H_{\rm dip}+\frac{4A_0}{\mu_0 M_0\Delta_0}\frac{\partial m}{\partial x}\right)+ \frac{K_S^0}{\mu_0 M_0}\sin(2\phi) \right]\, .\nonumber\\
\label{eq:1dmodel}
\end{eqnarray}
The second term on the RHS of Eq.~(\ref{eq:1dmodel}) is the entropic field as derived in  Eq.~(\ref{eq:ET})  while the first term is the TIDF. Both fields depend on the DW position $q$. 
The results of the 1D model calculations are plotted in Fig.~\ref{fig:Fig6}. For $d=2\sigma_L$ (Fig.~\ref{fig:Fig6}(a)) the model gives equal results with or without TIDF, as expected (the  TIDF is null in this case) and the agreement with simulations is good. For $d=4\sigma_L$ the model without TIDF (purple dashed line) predicts no DW motion, as expected from the ET since the DW is outside the temperature gradient. On the other hand, the model with the TIDF (black dotted line) predicts DW motion and shows a better agreement with simulations confirming our hypothesis.
By using the 1D model it is also possible to estimate the WB thermal gradient:
\begin{eqnarray}
\nabla T_{W}=\frac{\Delta_0\alpha_{\perp} K_S^0}{4A_0(\partial m/\partial T)}\, .
\label{eq:WB}
\end{eqnarray}
Due to the presence of $\partial m/\partial T$ the WB also depends on the absolute temperature $T$ which affects  $\partial m/\partial T$~\cite{Schlickeiser2014}. The WB as function of temperature is plotted in Fig.~\ref{fig:Fig6bis}. The blue points represent the maximum value of $\nabla T(x)$ (being a Gaussian profile, $\nabla T$ is not constant) applied in the simulations for different laser temperatures. The crossing of the WB occurs at $T\approx660K$, in reasonable agreement with our observation ($T=600K$, Fig.~\ref{fig:Fig3}(b)). The small difference could be given by the effect of the TIDF or by the uncertainty on the 1D parameters ($K_S^0$,$\Delta_0$).\footnote{$\Delta_0$ is calculated by fitting the static Bloch profile obtaining $\Delta_0=30\ {\rm nm}$. $K_S^0$ is obtained by calculating the static DW widths ($\Delta_1$,$\Delta_2$) and energies ($\epsilon_1$,$\epsilon_2$) for in-plane and out-of-plane DW ($\phi_1=0$, $\phi_2=\pi/2$) and using the relation $K_S^0=(1+ab)/a$ (Ref.~\cite{Hillebrands3}), with $a=(\Delta_1/\Delta_2)^2$ and $b=(\epsilon_1/\epsilon_2)^2$, obtaining $K_S^0\approx 2.9 \times 10^5\ {\rm J/m^3}$}
\r{Magnetostatic effects on thermally induced DW motion were already discussed by Berger~\cite{Berger1985} . Despite the common magnetostatic origin,  the TIDF shown here presents some differences: In Ref.~\cite{Berger1985} the thermal gradient affects the magnetostatic energy of the domains (much more relevant in bulk samples such as the ones analysed by Berger), whereas here, it is the Gaussian temperature profile by itself that generates new magnestostatic volume charges which give rise to the TIDF. }
\subsection{Magnonic spin transfer torque}
\label{subsec3}
\begin{figure}[H]

\includegraphics[width=0.45\textwidth]{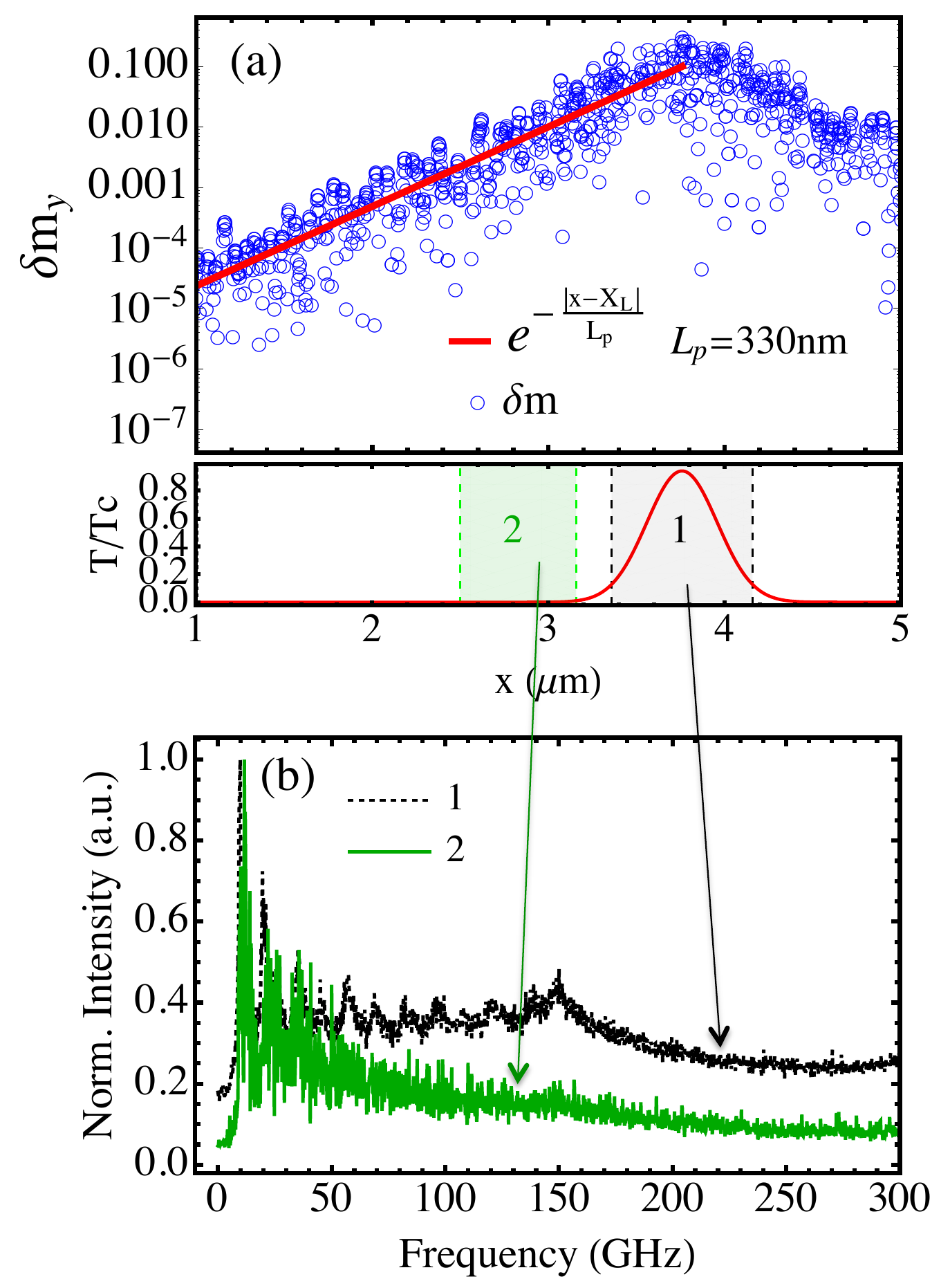}
\caption{(a) Magnon accumulation as defined in Ref.~\cite{Ritzmann2014} ($\delta m_y(x,t)=m_y(x,t)-m_y(x,0)$) for $X_L=3.76{\rm \mu m}$. \b{The time $t$ at which the magnons accumulations is calculated is $t=10$ns, long enough so that magnons have propagated along the strip and they have reached an equilibrium state.} Magnons decays exponentially as shown by the fit with $   e^{-|x-X_L|/L_p} $, where $L_p=330$nm is the magnon propagation length. (b) FFT intensity as function of magnons frequency in region 1: $X_L\pm 330$nm (below the laser spot, black dots) and region 2: $(X_L-5\sigma_L)\pm 330$nm (outside the laser spot, green line).} 
\label{fig:Fig7}
\end{figure}

\begin{figure}[H]
\centering
\includegraphics[width=0.4\textwidth]{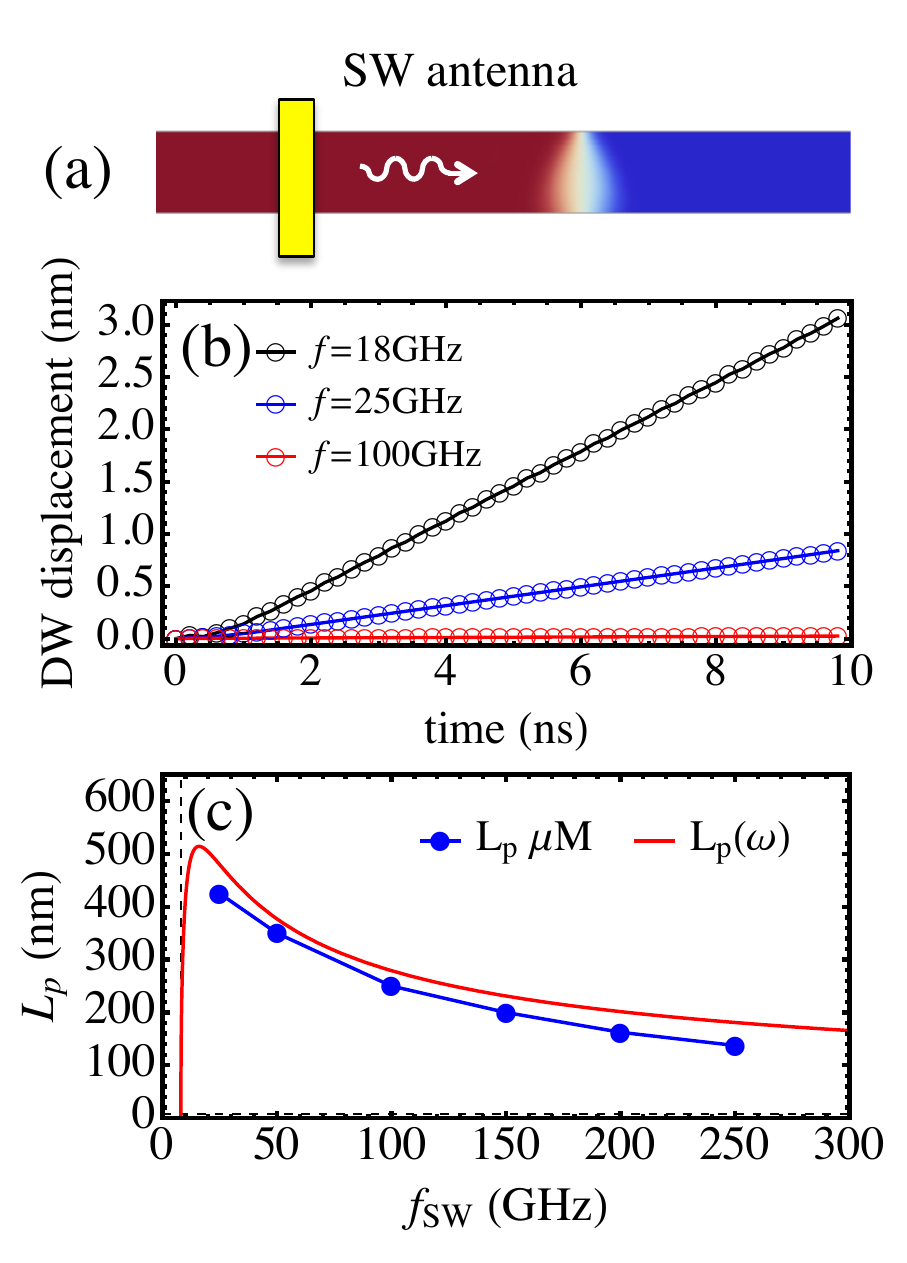}
\caption{(a) Schematic representation of the monochromatic spin waves simulations. (b) DW displacement as function of time for different frequencies. The maximum displacement is obtained for the lowest frequency ($f=18$ GHz). (c) SW propagation length as function of SW frequency. }
\label{fig:spinwaves}
\end{figure}

\begin{figure}[H]
\centering
\includegraphics[width=0.5\textwidth]{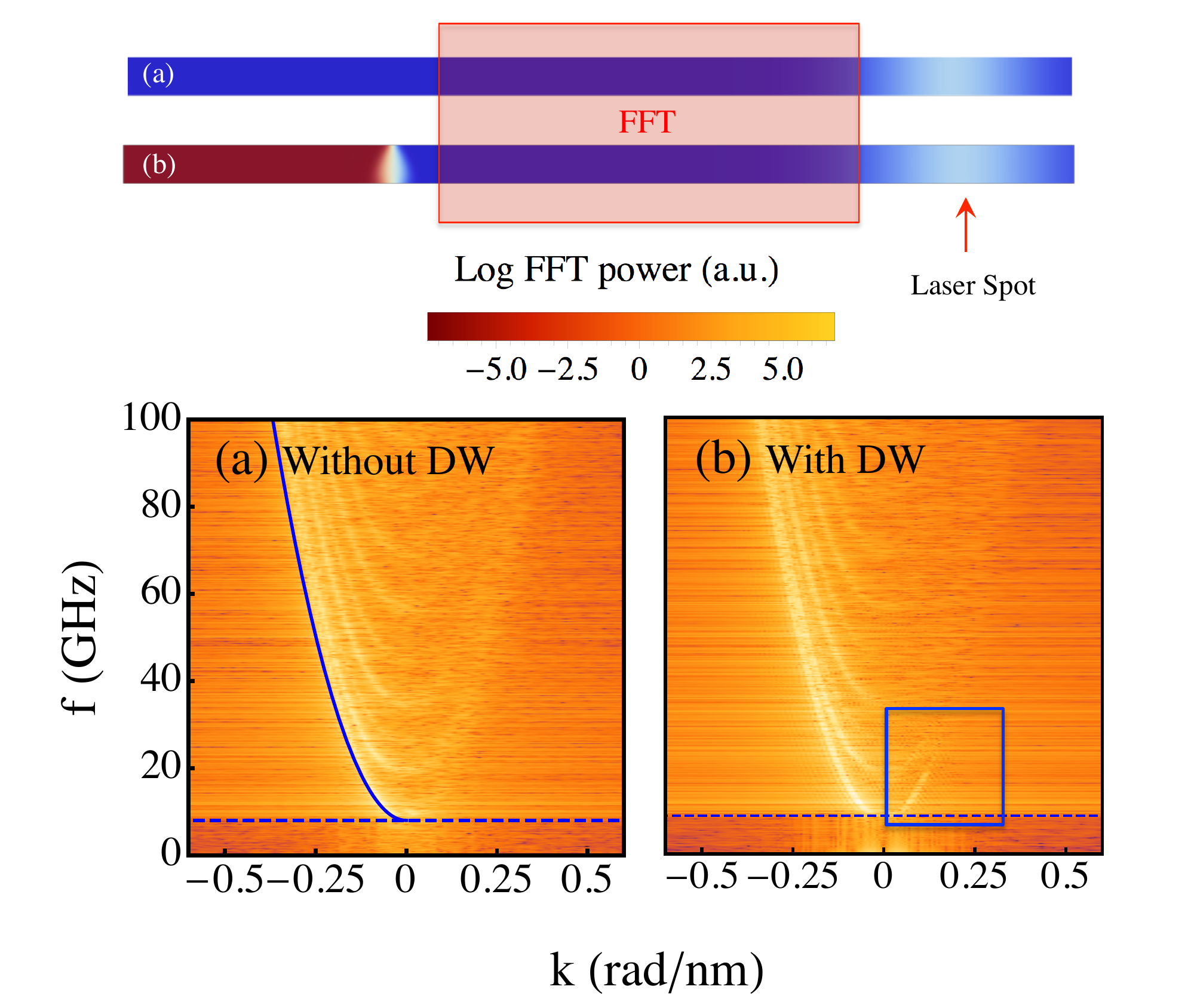}
\caption{Log FFT intensity as function of frequency $f$ and wave vector $k$ calculated at $X_{0}\pm 330$nm with (b) or without (a) the DW. The plot shows typical SW dispersion curves. The left branches correspond to magnons propagating from right to left (away from the laser spot),  while right branches correspond to  magnons propagating from left to right (towards the laser spot). A small right branch at low frequency can be observed in the case with DW (b), which corresponds to magnons reflected by the DW. Blue dashed line indicates the cut-off at $f\approx 9$GHz, \r{while the blue solid line in (a) corresponds to Eq.~\ref{eq:DR}, which shows a good agreement with the FFT intensity.}}
\label{fig:omegaVSk}
\end{figure}

Magnons can drive the DW either towards the hot or the cold part depending on their interaction with the DW: they drive the DW towards the cold part if they are reflected by the DW~\cite{Han2009,Wang2012,Yan2015} due to linear momentum transfer, while  they drive the DW towards the hot part if they pass through the DW due to angular momentum transfer~\cite{Yan2011,Wang2012,Kim2015b}. As already mentioned, we observe DW motion towards the hot part for $d=2,3\sigma_L$ (Fig.~\ref{fig:Fig2}(b)), DW motion towards the cold part for $d=4,5\sigma_L$ (Fig.~\ref{fig:Fig2}(e), inset of Fig.~\ref{fig:Fig3}(a)) and no DW motion for $d\geq 6\sigma_L$ (Fig.~\ref{fig:Fig2}(h), inset of Fig.~\ref{fig:Fig3}(a)).\\
At $d=2,3\sigma_L$ the motion is probably due to an averaged ET (as commented in Sec.~\ref{secIII}) and it is not possible to isolate the effect of magnons. For $d\geq 6\sigma_L$ the magnons are already damped and therefore they do not interact with the DW. In fact, by fitting the magnon accumulation~\cite{Ritzmann2014} $\delta m_y(x,t)=m_y(x,t)-m_y(x,0)$, we estimate a magnon propagation length $L_p=330$nm (Fig.~\ref{fig:Fig7}(a)). This means that at $d=4,5\sigma_L$ ($200$ nm and $400$ nm respectively from the end of the laser spot) the DW is within the magnon propagation length, while at $d=6\sigma_L$ the DW is at $\approx2L_p$, where magnons are clearly damped.  Therefore, at $d=4,5\sigma_L$, where $\nabla T(X_{DW})\approx 0$ the motion towards the cold part should be given by the $\mu STT$. To better understand such behavior, thermally excited magnons were analysed  by means of 2-Dimensional Fast Fourier Transform (FFT) in the middle of the strip ($y_0=40nm$) i.e. by calculating the FFT power~\cite{Han2009}
\begin{eqnarray}
\tilde{m_y}(\omega ,k_x)=\mathcal{F}_{\rm 2D}\left[m_y(x,y_0,t)-m_y(x,y_0,0)\right]\, ,
\end{eqnarray}
where the FFT is calculated with respect to $\{x,t\}$. Fig.~\ref{fig:Fig7}(b) shows the normalized  magnons frequency spectrum ($\sum_{k_x}\tilde{m_y}(\omega , k_x)$) at the laser spot (LS) (region 1: $X_L\pm 330$nm) and right before the LS (region 2: $(X_L-5\sigma_L)\pm 330$nm). At the LS (black dots) magnons have a wide range of frequency  while before the laser spot (green line) only low frequency magnons have propagated, in agreement with previous observation~\cite{Ritzmann2014}. The cut-off at $f_0\approx 9{\rm GHz}$ is due to lateral width confinement~\cite{Han2009}. Therefore, the average magnon propagation length, previously calculated (Fig.~\ref{fig:Fig7}(a)), is mainly related to low frequency magnons. This is a relevant observation since the magnons frequency strongly affects magnons transmission or reflection at the DW~\cite{Wang2012}. In particular, low frequency magnons are likely to be reflected~\cite{Wang2012,Han2009} and would produce motion towards the cold part.
To further understand the interaction between magnons and the DW, the DW dynamics excited by monochromatic spin waves  (SW) was analysed in the same Py strip. SW were locally excited by a transverse sinusoidal field $\HH_a(x)=H_0\sin(2\pi f)\hat{y}$ at a distance of \r{$100$nm} from the DW (Fig.~\ref{fig:spinwaves}(a)). The excitation region has dimensions $10\times80\times10\ {\rm nm}$ and $\mu_0H_0$ is set to \r{$10$mT}. The DW dynamics by different frequencies is shown in Fig.~\ref{fig:spinwaves}(b), while the spin wave propagation length as function of frequency is depicted in Fig.~\ref{fig:spinwaves}(c). Consistently with previous analysis~\cite{Han2009, Kim2012} the DW moves towards the cold part (in this case cold means away from the antenna position i.e. in the same direction as magnons propagation) for low frequency, $f=18,25$ GHz, while no motion towards the hot direction is observed within the maximum applied frequency, $f_{\rm max}=100$ GHz (Fig.~\ref{fig:spinwaves} (b)). Moreover,  the monochromatic analysis allows to study the frequency dependent magnon propagation length, and indeed it confirms that magnon propagation length decays with the magnons frequency (Fig.~\ref{fig:spinwaves}(c)). Note that the propagation length of low frequency magnons  is in good agreement with our calculation for thermal magnons ($L_p=330$ nm). \r{Furthermore, following Ref.~\cite{Ritzmann2014}, the frequency dependent propagation length $L_p(\omega)$ can be estimated as 
$1/(\alpha_{\perp}\omega) \partial\omega/\partial k $ and $\partial\omega/\partial$ k can be calculated from the spin waves dispersion relation in our system~\cite{Stancil2009}
\begin{eqnarray} 
\omega=\omega_0+l_{ex}^2\omega_Mk^2\, ,
\label{eq:DR}
\end{eqnarray}
where $l_{ex} $ is the exchange length and $\omega_M=\gamma_0M_0m_e(T)$. The cut-off frequency $\omega_0=2\pi f_0$ is taken from simulations (this expression for the spin waves dispersion relation  will be compared with the 2D FFT intensity in Fig.~\ref{fig:omegaVSk}(a) giving a good agreement). We finally obtain 
\begin{eqnarray}
L_p(\omega)=\frac{1}{\alpha_{\perp}\omega}\frac{\partial\omega}{\partial k}=\frac{2l_{\rm ex}}{\alpha_\perp}\frac{(\omega_M(\omega-\omega_0))^{1/2}}{\omega}\, .
\label{eq:lp}
\end{eqnarray}
Eq.~\ref{eq:lp} is also plotted in Fig.~\ref{fig:spinwaves} showing a good agreement with the simulation results. At high frequency, where $l_{ex}^2\omega_Mk^2\gg \omega_0$, Eq.~\ref{eq:lp} simply reduces to  $L_p\cong\lambda/(\pi\alpha)$, where $\lambda$ is the magnon wavelength.\cite{Ritzmann2014}}

In the laser spot case, an additional proof of magnons reflection is given by the FFT power in a region $X_{0}\pm 330$ nm, where $X_0$ is chosen in order to remain outside both the TG and the DW as sketched in Fig.~\ref{fig:omegaVSk}. The FFT is performed with (Fig.~\ref{fig:omegaVSk} (b)) or without  (Fig.~\ref{fig:omegaVSk}(a)) the DW. The left bright branches correspond to magnons propagating from right to left, moving away from the laser spot, as expected. In the FFT with the DW (Fig.~\ref{fig:omegaVSk}(b)) a small branch appears on the right side, which corresponds to magnons propagating from left to right, towards the laser spot, as a consequence of reflection  by the DW~\cite{Han2009}. Therefore we conclude that the DW motion towards the cold part is due to low frequency magnons, excited by the laser, which have larger propagation length and are reflected by the DW. The result is different from that predicted by Kim and Tserkovnyak~\cite{Kim2015b}, where DW was supposed to move towards the hotter region due to magnons transmission through the DW. \r{The difference is probably due to the different magnon wavelength: in Ref.~\cite{Kim2015b} the authors assumed that the thermal magnon wavelength is much shorter than the DW width, focusing on magnon transmission and the adiabatic STT.  In our case, low-frequency (large-wavelength) modes dominate outside the TG due to their larger propagation length and they are mainly reflected by the DW.} \t{From Eq.~\ref{eq:DR} we can also estimate the wavelength of the reflected magnons. The frequency  range of the reflected branch in Fig.~\ref{fig:omegaVSk}(b) is approximatively $f_0<f<20$GHz which corresponds to a wavelength range $50<\lambda<400$nm, larger than or comparable to the DW width parameter $\Delta_0=30$nm}. \r{ Inside the TG ($d=2\sigma_L$) the motion is towards the hotter region, consistent with the result of Kim and Tserkovnyak.~\cite{Kim2015b}}
In the \full simulations the DW moves towards the hot part for $d=4,5\sigma_L$ meaning that the dipolar field is stronger than the $\mu$STT in this system. 
\t{As commented in Sec.~\ref{secII}, our analysis of the magnonic STT neglects magnons with $\lambda<5$nm. Due to their small propagation length ($L_p\sim80$nm) they can have an effect only at $d=2,3\sigma_L$, when the DW is inside the TG.  Since their wavelength is much smaller than the DW width, they are expected to pass adiabatically through the DW, moving it towards the hotter region as the ET. This would lead to higher velocities for the \magnonic case at $d=2,3\sigma_L$, however, their contribution is expected to be small since their propagation length is comparable to the full DW width ($\pi\Delta_0\sim90$nm) and therefore, angular momentum transfer is strongly reduced.  }

\subsection{Realistic sample}
\label{subsec4}
\begin{figure}[H]
\centering
\includegraphics[width=0.42\textwidth]{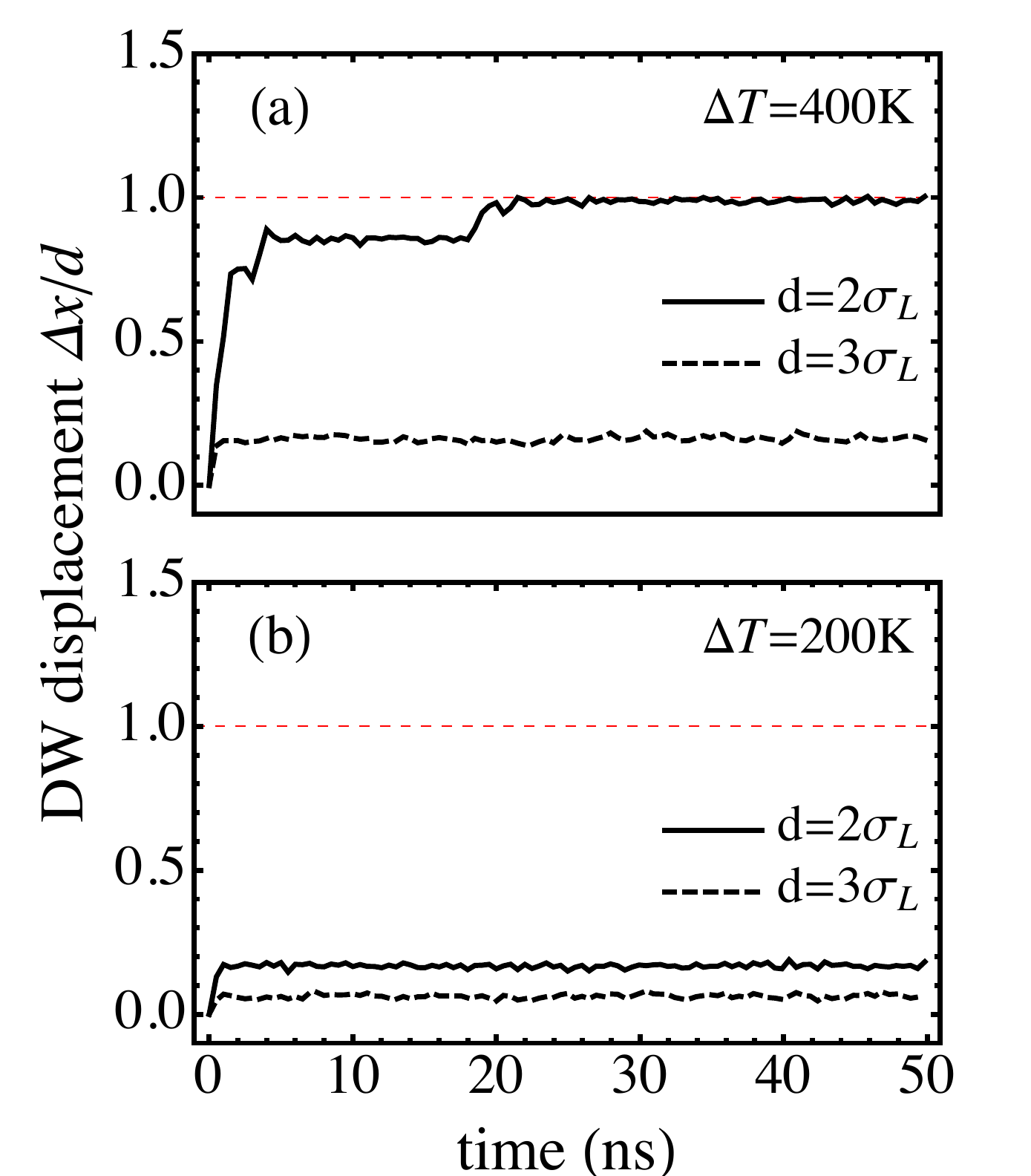}
\caption{DW displacement as funciton of time for $\Delta T=400$K (a) and $\Delta T=200$ K (b), for $d=2\sigma_L$ (full black line) and $d=3\sigma_L$ (black dashed line). The DW reaches the laser spot (red dashed line at $\Delta x/d=1$) only for $\Delta T=400$K at $d=2\sigma_L$.}
\label{fig:Fig9}
\end{figure}

Our previous results, as well as former theoretical investigations~\cite{Schlickeiser2014,Kim2015b,Yan2015}, were obtained for a perfect strip where even a small driving force was able to move the DW. However, it is well known that defects or inhomogeneities give rise to DW pinning and a finite propagation field ($H_p\neq0$) below which the DW remains pinned. We have also analysed DW motion by TG under realistic conditions to see in which case the TG is strong enough to depin the DW.  The introduction of  edge roughness with a characteristic size of $2.5$ nm  gives rise to a DW propagation field of $\mu_0H_p=(3.5\pm0.5)$mT. Also in this case the sample temperature follows Eq.~(\ref{eq:Tprofile}) but the strip temperature $T_0$ is set $T_0=300$K as it would be in conventional experiments. Considering the same $\Delta T$ of the previous analysis and in order to remain below $T_C$ we can only apply $\Delta T=400$K and $\Delta T=200$K.
As shown in Fig.~\ref{fig:Fig9} and in the corresponding movie~\cite{Note4}, the DW moves towards the laser spot only if it is close enough to the laser spot ($d\leq2\sigma_L$) and only if $\Delta T=400$K. Therefore, under realistic conditions, long-range dipolar field and $\mu$STT are not strong enough to move the DW as they are likely hindered below the propagation field in typical experiments. This observation is indeed in agreement with recent experimental observation~\cite{Tetienne2014a} where the DW motion towards the (close) laser spot was  succesfully explained by the sole effect of ET~\cite{Tetienne2014a}.
\section{Conclusions}
\label{sec4}
DW motion by Gaussian temperature profiles was analysed in a Py strip under perfect and realistic conditions. Apart from the already known entropic and magnonic contributions, a third driving force was observed due to a thermally induced dipolar field. Such force drives the DW towards the hotter region. An expression for the entropic field was derived in terms of  the DW energy and  compared with previous expressions showing equal results. The entropic torque pushes the DW towards the hot part and dominates the DW dynamic when the DW is within the TG, while the dipolar field dominates when the DW is outside the TG. In fact, the $\mu$STT drives the DW towards the cold part due to the prevalence of low frequency magnons, which propagate over larger distances ($L_p\approx330$nm) and are reflected by the DW in the studied sample. Finally, under realistic conditions,  the entropic torque is strong enough to move the DW only if the laser spot is closer than $2\sigma_L$ and $\Delta T\geq400$K. \r{These conclusions can be generalized to other in-plane samples, but we cannot rule out that, in systems with low damping the magnonic STT could overcome the thermally induced dipolar field outside the TG. } These results give important insights into the different mechanism responsible for DW motion under thermal gradients  and allows for comparison with experimental results in these  systems. 

\section*{Acknowledgement}
S.M. would like to thank M. Voto and R. Yanes-Diaz for useful discussions. This work was supported by Project WALL, FP7-
PEOPLE-2013-ITN 608031 from the European Commission, Project No. MAT2014-52477-C5-4-P from the Spanish government, and Project No. SA282U14 and SA090U16 from the Junta de Castilla y Leon.


%
%

\bibliographystyle{apsrev4-1}
\bibliography{library}

\end{document}